\documentclass{optica-article}

\journal{opticajournal}

\articletype{Research Article}

\usepackage{lineno}
\usepackage{bm}

\begin{document}

\title{Generation of 12~dB squeezed light from a waveguide optical parametric amplifier using a machine-learning-controlled spatial light modulator}

\author{Gyeongmin Ha,\authormark{1} Kazuki Hirota,\authormark{1} Takahiro Kashiwazaki,\authormark{2} Takumi Suzuki,\authormark{1} Akito Kawasaki,\authormark{3,4} Warit Asavanant,\authormark{3,4} Mamoru Endo,\authormark{1,3} and Akira Furusawa\authormark{1,3,4*}}

\address{\authormark{1}Department of Applied Physics, School of Engineering, The University of Tokyo, 7-3-1 Hongo, Bunkyo-ku, Tokyo 113-8656, Japan\\
\authormark{2}Device Technology Labs, NTT, Inc., 3-1 Morinosato Wakamiya, Atsugi, Kanagawa 243-0198, Japan\\
\authormark{3}Optical Quantum Computing Research Team, RIKEN Center for Quantum Computing, 2-1 Hirosawa, Wako, Saitama 351-0198, Japan\\
\authormark{4}OptQC Corp., 1-21-7 Nishi-Ikebukuro, Toshima-ku, Tokyo 171-0021, Japan
}

\email{\authormark{*}akiraf@ap.t.u-tokyo.ac.jp}
\begin{abstract*} 
We demonstrate the generation of $12.1 \pm 0.2$~dB squeezed light from a periodically poled lithium niobate (PPLN) waveguide optical parametric amplifier (OPA). While single-pass OPAs offer squeezed light with THz-order bandwidths, loss from spatial mode mismatch between the squeezed light and the local oscillator (LO) previously capped the squeezing level at $\sim$10~dB [K. Hirota et al., Opt. Express 34, 7958 (2026)]. In this work, we minimize this loss by introducing a machine-learning-optimized spatial light modulator (SLM) in the path of the LO. Specifically, we employed a double-reflection configuration to increase the spatial degrees of freedom, and directly used the measured squeezing level as the optimization's objective function.
\end{abstract*}

\section{Introduction}
Squeezed light is an essential resource for many applications, including quantum metrology \cite{Aasi.et.al.2013LIGO, Ganapathy.et.al.2023LIGO} and quantum information processing \cite{Takeda_and_Furusawa_2019_toward_FTUPQC}. Over the years, the theoretical and experimental foundations for optical quantum computers have been laid out \cite{Yokoyama_teleportation_based_QC_2013, warit_2D_cluster, Takeda_and_Furusawa_2019_toward_FTUPQC}. In this scheme of teleportation-based quantum computing, the resource state, called cluster state, is generated using squeezed light.

Squeezed vacuum states with higher squeezing levels are essential for improving computational accuracy in teleportation-based quantum computing. Furthermore, the clock frequency of an optical quantum computer is fundamentally limited by the bandwidth of the squeezed light. Therefore, high-speed, low-noise information processing requires a squeezed light source with high squeezing level and a broad bandwidth. Historically, optical parametric oscillators (OPOs) have been employed to generate squeezed states \cite{9dB_2007, squeezing_at_1550nm_13.5dB, 15dB_OPO_PhysRevLett.117.110801}. 
OPOs typically generate, but are not limited to, squeezed states in a single mode \cite{Roh_Robustsqueezedlight_2021, Higherordersqueeze_2022}. In both cases, the spatial modes of the squeezed states are well-defined by the resonator geometry \cite{LaserBeams_and_Resonators}. However, this same resonator structure limits the bandwidth of the squeezed light to the MHz up to GHz range \cite{Ast_Broadband_OPO_1GHz:13}. On the other hand, single-pass waveguide optical parametric amplifiers (OPAs) offer a much simpler, cavity-free experimental implementation. Furthermore, being limited only by their phase matching, OPAs inherently provide THz-order bandwidths.

Given the ability of waveguide OPAs to generate broadband squeezed states, various efforts have been made to increase the squeezing level of squeezed states generated from waveguide OPAs \cite{Hirano_3.4dB_2007, Hirano_5dB_2011, Kashiwazaki_8.3dB_2023}. However, the squeezing levels remained far below those of OPOs, with the intrinsic loss of the waveguide and loss from mode mismatch being the main limiting factors. In recent years, a low-loss periodically poled lithium niobate (PPLN) waveguide OPA was fabricated, with propagation loss of less than 0.1~dB/cm \cite{Kashiwazaki_6THz_0.1dBpercm}. This led to the generation of a squeezed state with bandwidth as wide as 6~THz \cite{Kashiwazaki_6THz_0.1dBpercm} and squeezing level as high as $10.1 \pm 0.2$~dB \cite{Hirota10dB:26}.

Various approaches have been taken to minimize the loss from mode mismatch, such as using an output from a waveguide as the local oscillator (LO) \cite{Hirano_3.4dB_2007, Kashiwazaki_6dB_2.5THz_2020}, and using a spatial light modulator (SLM) to improve mode matching \cite{Amari:23}. In these works, as direct measurement of the visibility between the squeezed light and the LO is difficult, the visibility between the transmitted light at the fundamental wavelength (referred to as the `probe beam' in this paper) and the LO was used as a proxy. However, Hirota \textit{et al}. (2026) identified that this indirect measure can be misleading, reporting that while a visibility of 99\% was achieved between the probe beam and the LO, the loss due to mode mismatch was estimated to be 4\% \cite{Hirota10dB:26}. In the same work, the authors stated that the primary factor degrading the squeezing level was the spatial mode mismatch between the squeezed light and the LO, accounting for 4\% out of the total 8\% loss.

In this work, we introduce an SLM into the LO path to minimize this loss from mode mismatch, adopting a double-reflection configuration where the LO is reflected twice off the SLM surface. This configuration significantly increases the degrees of freedom for spatial mode control compared to single-reflection setups. Crucially, rather than relying on the probe beam visibility as a proxy, we directly utilize the squeezing level itself as the objective function for the machine learning optimization. This approach allows us to surpass the limits of previous SLM-based optimizations, which reported a mode-matching efficiency of 95\% (corresponding to $\sim$5\% loss) \cite{Amari:23}. By directly optimizing the final figure of merit, we reduced the loss attributed to mode mismatch to approximately 0.4\%, resulting in a total system loss of only 4.4\% and the successful generation of $12.1 \pm 0.2$~dB squeezed light.

\section{Experimental Setup}
\begin{figure}[htbp]
    \centering\includegraphics[width=13.2cm]{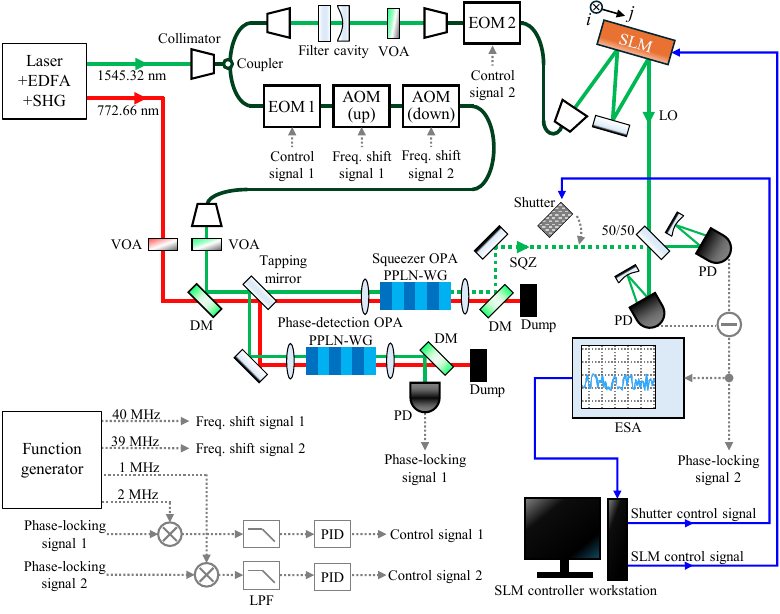}
    \caption{Schematic of the experimental setup used to optimize phase mask of spatial light modulator (SLM) and the squeezing level measurement. Green and red lines represent fundamental and second harmonic beams (1545.32~nm and 772.66~nm) in free space. Dark green is the fundamental beam inside an optical fiber. Gray dotted lines represent the electrical signals used for phase control feedback, and blue solid lines represent control signals for phase mask optimization. The squeezed vacuum states of light (SQZ) were generated in a periodically poled lithium niobate waveguide optical parametric amplifier (squeezer OPA PPLN-WG). A tapping mirror and a separate waveguide (phase-detection OPA PPLN-WG) was used to generate phase-locking signal \cite{Hirota10dB:26}. A SLM was placed in the path of the local oscillator (LO) to minimize loss from spatial mode mismatch between the squeezed light and the LO. Reflecting the LO twice off the SLM provided additional degrees of freedom for spatial mode control.
    EDFA: erbium doped fiber amplifier; SHG: second harmonic generator; EOM: electro-optic modulator; AOM: acousto-optic modulator; VOA: variable optical attenuator; DM: dichroic mirror; PD: photodiode; LPF: low pass filter; PID: proportional-integral-derivative controller; ESA: electrical spectrum analyzer.}
    \label{fig:SLM_experimental_setup}
\end{figure}

Figure~\ref{fig:SLM_experimental_setup} shows the setup used in this work. The squeezed light was generated and the phase lock was performed in the same manner as \cite{Hirota10dB:26}, using PPLN waveguide OPAs fabricated in the same method as \cite{Kashiwazaki_6THz_0.1dBpercm}.
A seed laser system (NKT Photonics, Koheras Harmonik) delivered free-space output at the fundamental wavelength of 1545.32~nm and its second harmonic at 772.66~nm. The fundamental beam was coupled into a fiber and split in half by a fiber coupler to be used as the probe beam and the LO. The phase of the probe beam was shifted by an electro-optic modulator (EOM 1). The probe beam was also frequency-shifted by $+1$~MHz using two acousto-optic modulators (AOM (up) and AOM (down)) (Chongqing Smart Science \& Technology Development, SGTF-40-1550-1P).

Using a dichroic mirror, this frequency-shifted probe beam was combined with the pump beam (772.66~nm). A portion of the pump and probe beams was tapped using a tapping mirror and injected into the phase-detection OPA using an aspherical lens with a focal length of 4.5 mm. Another aspherical lens with the same focal length was used to collimate the transmitted light in free space. In the phase-detection OPA, the probe beam underwent parametric amplification, resulting in an output beam whose power oscillates at 2~MHz. This oscillation was detected using an InGaAs photodiode, whose electrical output signal was passed through a mixer and a low pass filter (LPF) to produce an error signal. This error signal was fed into a PID controller to produce a control signal, which was fed back to EOM 1 to lock the relative phase between the probe beam and the pump beam. The pump beam and probe beam transmitted by the tapping mirror were coupled into the squeezer OPA and back into free space using two aspherical lenses, each with a focal length of 4.5 mm. This produced a parametrically-amplified probe beam and squeezed light. The pump beam was removed by a dichroic mirror. An optical shutter was placed in the path of the parametrically-amplified probe beam and squeezed light for automatic squeezing level measurement used in the machine learning optimization loop.

The LO beam was collimated into free space and passed through a filter cavity to suppress noise from amplified spontaneous emission (ASE). This was once again coupled back into a fiber, where an EOM (EOM 2) was used to shift the phase, and coupled to free space. The LO underwent two successive reflections from a spatial light modulator (Santec, SLM-200) (SLM) partitioned into two halves. This double-reflection configuration allowed for greater degrees of freedom compared to a single reflection \cite{Amari:23}. The Gaussian beam radii were $1.2$~mm, and the angles of incidence were less than 7 degrees for both incident beams. The SLM and the optical shutter in the path of the squeezed light were controlled by a workstation (SLM controller workstation). The parametrically amplified probe beam and the squeezed light co-propagated from the squeezer OPA and were interfered with the LO at a 50/50 beam splitter. For balanced homodyne detection, we utilized two InGaAs photodiodes (Laser Components, IGHQEX0500-1550-10-1.0-SPAR-TH-40) with photoelectric quantum efficiency of around 99\% and an in-house electrical circuit. Two curved mirrors were used as retroreflectors \cite{15dB_OPO_PhysRevLett.117.110801}. The output of the homodyne detector was sent to an electrical spectrum analyzer (Keysight, N9010B) (ESA) and a phase lock loop. This feedback loop consisted of a mixer, LPF for error signal generation, followed by a PID controller to produce a control signal. This signal was fed back into EOM 2.

\section{SLM Phase Mask Generation}
In this section, we describe the steps taken to find the optimal phase mask of the SLM. A typical SLM has $H \times W$ number of pixels, with each pixel having an $k$-bit integer value corresponding to phase retardation of $0$ to $2 \pi$. The phase mask of an SLM can be determined by a $H \times W$ matrix $M$, with each matrix element $M_{i,j}$ having an integer value between 0 and $2^{k}-1$.
\begin{equation}
    M = 
    \begin{pmatrix}
        M_{1,1} & \cdots & M_{1,W} \\
        \vdots & \ddots & \vdots \\
        M_{H, 1} & \cdots & M_{H, W}
    \end{pmatrix},
    \qquad M_{i,j} \in \{0,1,\dots,2^{k}-1\}
\end{equation}
The SLM used in this work had an effective area size of $9.60~\mathrm{mm} \times 15.36~\mathrm{mm}$, a pixel resolution of $1200 \times 1920$, a pixel size and pitch of $7.8~\mathrm{\mu m}$ and $8~\mathrm{\mu m}$, respectively, and a 10-bit gray level per pixel. Therefore, the phase mask of the SLM used in this work can be determined by a $1200 \times 1920$ matrix, with each matrix element having an integer value between 0 and $2^{10}-1$.

As there are more than two billion possible phase masks, testing all possibilities is computationally impractical. To address this, previous implementations of machine-learning-controlled SLMs for homodyne mode-matching, such as Ref. \cite{Amari:23}, utilized a single-reflection configuration and restricted the phase mask parameterization to simple Gaussian profiles. However, such simple profiles are fundamentally limited to correcting basic wavefront curvatures and cannot adequately compensate for complex spatial aberrations, astigmatism, or higher-order mode deformations. In this work, we implemented a double-reflection configuration and a completely new set of parameters by parameterizing the masks as a combination of Zernike polynomials and two perpendicular cylindrical lenses. This parameter set, represented by parameter vector $\bm{\beta}$, was optimized using a Bayesian optimization (BO) algorithm. The gray-level assignment of the phase mask was represented by matrix $M(\bm{\beta})$ whose elements were defined as the sum of two terms: $\{\Psi(\bm{\beta})\}_{i,j}$ and $\mathcal{W}_{i,j}$.
\begin{equation}
    \{M(\bm{\beta})\}_{i,j} = \left\lfloor \left( \{\Psi(\bm{\beta})\}_{i,j} + \mathcal{W}_{i,j} \right) \pmod{2^{k}} \right\rfloor
\end{equation}
Here, $\Psi(\bm{\beta})$ is the phase mask matrix generated using parameter vector $\bm{\beta}$, and $\mathcal{W}$ is the starting phase mask matrix.

The elements of the phase mask, $\{\Psi(\bm{\beta})\}_{i,j}$, are defined piecewise for the left ($L$) and right ($R$) half-panels, where a normalized coordinate system $(\xi,\eta)$ with $\xi,\eta \in [-1,1]$ is used.
\begin{equation}
    \{\Psi(\bm{\beta})\}_{i,j} = 
    \begin{cases} 
        \Phi \left( \xi_L(j), \eta(i) \mid \bm{\beta}_L \right) \qquad\text{for} &1 \le j \le W_{\text{half}} \\
        \Phi \left( \xi_R(j), \eta(i) \mid \bm{\beta}_R \right) \qquad\text{for} &W_{\text{half}} +1 \le j \le W
    \end{cases}
\end{equation}
Here, $W_{\text{half}}= W/2$ is the half-width of the SLM. $\bm{\beta}_L$ and $\bm{\beta}_R$ are the parameter vectors for each half-panel, with $\bm{\beta} = \begin{pmatrix}
    \bm{\beta}_L &\bm{\beta}_R
\end{pmatrix}$.
The normalized coordinates are given by:
\begin{align}
    \eta(i) &= \frac{2(i - 1)}{H - 1} - 1, \quad \eta \in [-1, 1] \\
    \xi_L(j) &= \frac{2(j - 1)}{W_{\text{half}} - 1} - 1, \quad \xi_L \in [-1, 1], \quad \text{for } 1 \le j \le W_{\text{half}}\\
    \xi_R(j) &= \frac{2(j - W_{\text{half}} - 1)}{W_{\text{half}} - 1} - 1, \quad \xi_R \in [-1, 1], \quad \text{for } W_{\text{half}} +1 \le j \le W.
\end{align}
The scalar phase function $\Phi \left( \xi, \eta \mid \bm{\beta} \right)$ comprises a Zernike polynomial \cite{ZERNIKE1934689} expansion term $\Phi_{\text{Zernike}}(\rho, \phi)$ for wavefront correction and a generalized cylindrical lens term $\Phi_{\text{Lens}}(x_c, y_c)$ for mode-matching:
\begin{equation}
    \Phi(\xi, \eta \mid \bm{\beta}) = \Phi_{\text{Zernike}}(\rho, \phi) + \Phi_{\text{Lens}}(x_c, y_c)
\end{equation}
Here, $(\rho, \phi)$ are polar coordinates derived from the normalized Cartesian coordinates $(\xi, \eta)$, where $\rho = \sqrt{\xi^2 + \eta^2}$ and $\phi = \arctan(\eta / \xi)$. Due to the independent normalization of $\xi$ and $\eta$, the condition $\rho=1$ defines an ellipse inscribed within the rectangular half-panel boundaries.

Zernike polynomials $Z_n^m(\rho, \phi)$ from the 1st to the 5th radial order were used. The $\Phi_{\text{Zernike}}(\rho, \phi)$ was expressed as the linear combination of Zernike polynomials:
\begin{equation}
    \Phi_{\text{Zernike}}(\rho, \phi) = \sum_{n,m} C_{n}^{m} Z_n^m(\rho, \phi),
\end{equation}
where $C_n^m$ are the expansion coefficients corresponding to the Zernike polynomial of radial order $n$ and azimuthal order $m$ ($1\leq n\leq5$).

Unlike the Zernike terms, the cylindrical lens term $\Phi_{\text{Lens}}$ was calculated using physical pixel coordinates $(x_c, y_c)$ relative to the geometric center of each sub-aperture. These were defined as:
\begin{align}
    y_c(i) &= i - \frac{H+1}{2} \\
    x_c(j) &= 
    \begin{cases} 
        j - \frac{W_{\text{half}}+1}{2} \quad &\text{for} \quad 1 \le j \le W_{\text{half}} \\
        (j - W_{\text{half}}) - \frac{W_{\text{half}}+1}{2} \quad&\text{for} \quad W_{\text{half}} +1 \le j \le W
    \end{cases}
\end{align}
To allow for arbitrary astigmatic axis orientation, these coordinates are rotated by an angle $\theta_{\text{lens}} \in [0,\pi/2]$ and decentered by $(\mu_x, \mu_y)$ (in pixels):

\begin{equation}
    \begin{pmatrix} u \\ v \end{pmatrix} = 
    \begin{pmatrix} \cos\theta_{\text{lens}} & -\sin\theta_{\text{lens}} \\ \sin\theta_{\text{lens}} & \cos\theta_{\text{lens}} \end{pmatrix} 
    \begin{pmatrix} x_c - \mu_x \\ y_c - \mu_y \end{pmatrix}
\end{equation}
The final lens phase is applied as a quadratic function of the rotated coordinates:

\begin{equation}
    \Phi_{\text{Lens}} = \gamma_1 u^2 + \gamma_2 v^2
\end{equation}
where $\gamma_1$ and $\gamma_2$ are the lens strength parameters. 

In total, 40 parameters were optimized, represented by the vector $\bm{\beta} = \begin{pmatrix}
    \bm{\beta}_L &\bm{\beta}_R
\end{pmatrix}$, where the sub-vector for each side ($S \in \{L, R\}$) is:
\begin{equation}
    \bm{\beta}_S = \left[ C_{1,S}^{-1}, C_{1,S}^{1}, \dots, C_{5,S}^{5}, \gamma_{1,S}, \gamma_{2,S}, \theta_{\text{lens},S}, \mu_{x,S}, \mu_{y,S} \right].
\end{equation}

A Bayesian optimization (BO) algorithm was written in Python to find the optimal parameter vector $\bm{\beta}$ that produced the best squeezing level. This algorithm was run on an SLM controller workstation. Unlike conventional approaches that optimize proxy metrics such as interference visibility between the probe beam and the LO \cite{Hirano_3.4dB_2007, Kashiwazaki_6dB_2.5THz_2020, Amari:23}, this algorithm directly utilized the measured squeezing level (dB) as the objective function, ensuring the optimization targeted the final figure of merit.

To ensure the squeezing measurement was robust against power fluctuations in the LO, we implemented a differential measurement protocol. For each candidate phase mask, we recorded both the squeezed noise power ($P_{\text{sqz}}$, in dBm) and the shot-noise power ($P_{\text{shot}}$, in dBm) to obtain the squeezing level of $(P_{\text{shot}} - P_{\text{sqz}})$~dB. The shot-noise was measured by momentarily closing the signal arm shutter, leaving only the LO. This step compensated for variations in LO power reaching the homodyne detector caused by changes in SLM phase masks. Spectral data was acquired using an electrical spectrum analyzer (ESA) in zero-span mode (center frequency: 3~MHz, RBW: 1~MHz, VBW: 200~Hz), where a single squeezed noise (or shot-noise) measurement was an average of 100 consecutive sweeps for an accurate measurement.

This optimization used the BoTorch framework \cite{BoTorch2020}, which utilized a Gaussian Process (GP) regression model implemented via GPyTorch \cite{GPyTorch2018} as a surrogate to map the relationship between the parameter vector $\bm{\beta}$ and the measured squeezing level. The GP covariance was modeled using a Matern-5/2 kernel with Automatic Relevance Determination (ARD). To ensure robust hyperparameter fitting, the input parameter space was normalized to the unit hypercube, and the measured squeezing values were standardized (zero mean, unit variance) prior to model training. 

The optimization was initialized by evaluating a set of random points (20-40 points) equal to 10\% of the total iteration budget (200-400 iterations), ensuring sufficient coverage of the parameter space before the model-based acquisition began. In subsequent iterations, the next set of experimental parameters was selected using the q-Max-value Entropy Search (qMES) acquisition function \cite{pmlr-v70-wang17e}. This strategy selects points that maximize the mutual information between the observation and the global maximum, effectively balancing the exploration of the parameter space with the exploitation of high-squeezing regions.

The steps for obtaining the optimal phase mask of the SLM were as follows:
\begin{enumerate}
\item Using diagnostic phase patterns and observing the reflected beam's power distribution, the two incident beams of the SLM were steered to the center of each half-panel.
\item With no phase modulation on the SLM (i.e., $\mathcal{W}_{i,j}=0$ for all $i,j$), the squeezed light and the LO were manually aligned and the relative phases between the two beams were locked.
\item The Bayesian optimization algorithm was executed. Each optimization run consisted of between 200 and 400 iterations. At each iteration, the shot-noise (squeezed light blocked by the optical shutter) and the squeezed noise (both LO and squeezed light present) were measured at the homodyne detector using an electrical spectrum analyzer. The squeezing level was calculated from the shot-noise and the squeezed noise levels, and this value was used directly as the objective function to be optimized by the BO algorithm.
\item After each optimization run, the phase mask yielding the highest squeezing level $M(\bm{\beta}_{\text{optimal}})$ was displayed on the SLM, and the LO was manually realigned to correct for slight beam deviations caused by thermal expansion of optical components. 
\item The new optimal phase mask was defined as a new starting phase mask $\mathcal{W} \leftarrow M(\bm{\beta}_{\text{optimal}})$.
\item Using this updated starting phase mask $\mathcal{W}$, steps 3–5 were repeated.
\end{enumerate}

\section{Results}
\begin{figure}[htbp]
    \centering\includegraphics[width=13.5cm]{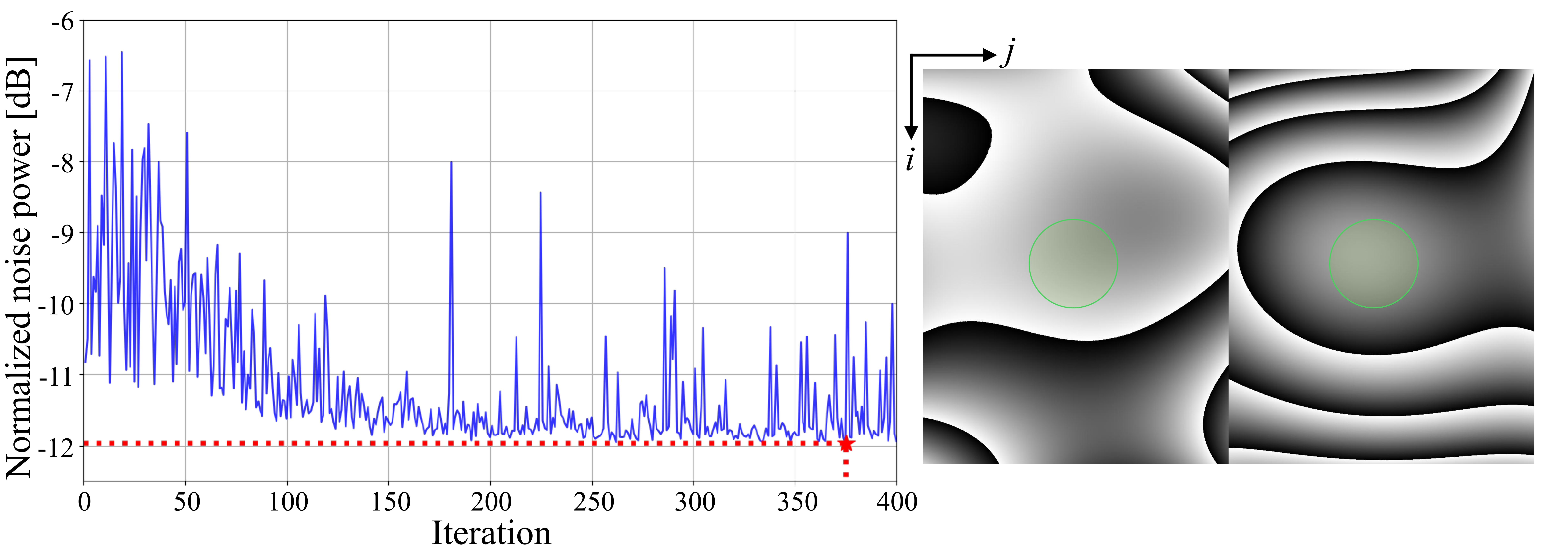}
    \caption{Left: Result of the last optimization run. At every iteration, the squeezing level was automatically measured and used as the objective function for the BO algorithm to optimize. The parameters tested in the first 40 iterations were randomly sampled. The best pattern found was at the 375th iteration (marked in red), initially yielding 12.0~dB. After careful re-alignment, the same pattern generated $12.1 \pm 0.2$~dB. Right: Visualization of this optimal phase mask, where the grayscale gradient represents the 10-bit phase modulation values (black $= 0$, white $= 1023$). The semi-transparent green circles indicate the Gaussian beam size (radii of 1.2~mm) for the two incident beams.}
    \label{fig:convergence_plot}
\end{figure}

Figure \ref{fig:convergence_plot} shows the squeezing level for 400 iterations in the last optimization run. Including the time taken for the squeezing level measurement, the first iterations typically took 4 seconds, which increased to around 15 
seconds on the 400th iteration.

\begin{figure}[htbp]
    \centering\includegraphics[width=13.5cm]{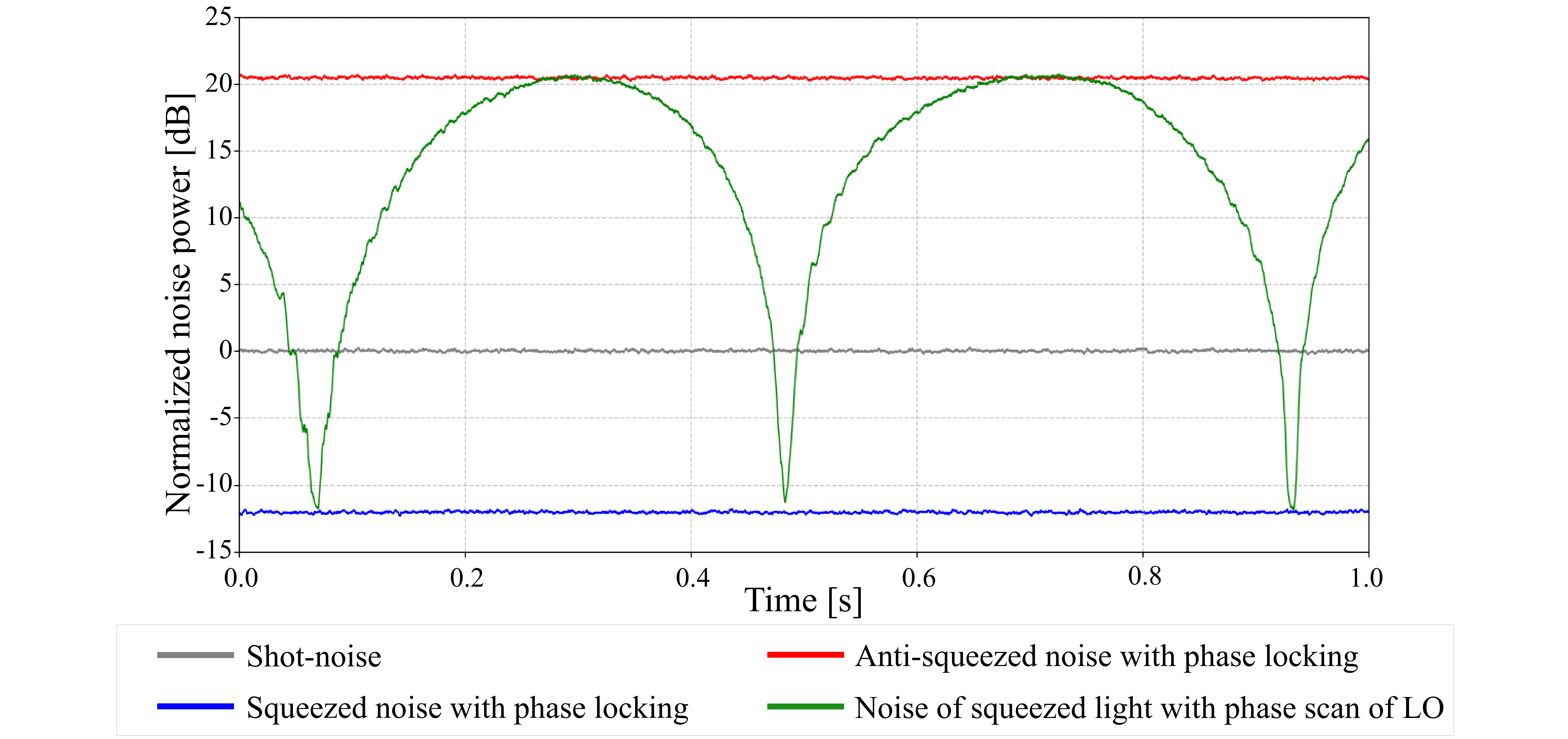}
    \caption{Various noise levels measured by balanced homodyne detection using an electrical spectrum analyzer operated in zero-span mode at a sideband frequency of 3~MHz, with a resolution bandwidth of 1~MHz and a video bandwidth of 100~Hz. The pump power was 585~mW, measured immediately after the squeezer OPA. The gray line represents the shot-noise level. The blue and red lines show the squeezed and anti-squeezed noise, respectively, normalized to the shot-noise level. Green line represents the normalized noise level of the squeezed light with the phase of the LO scanned.}
    \label{fig:12dB_zerospan}
\end{figure}

Following the steps mentioned above, we were able to increase the squeezing level from less than 9~dB and surpass the previous record of $10.1 \pm 0.2$~dB \cite{Hirota10dB:26}.
Figure \ref{fig:12dB_zerospan} shows the squeezed and anti-squeezed noise normalized to shot-noise for a pump power of 585~mW immediately after the squeezer OPA. This measurement was taken by an electrical spectrum analyzer operated in zero-span mode at a sideband frequency of 3~MHz, with a resolution bandwidth of 1~MHz and a video bandwidth of 100~Hz. The LO power was 16.5~mW. The clearance between the electrical circuit noise and the shot-noise level was greater than 28~dB at 3~MHz sideband frequency. A squeezing level of $12.1 \pm 0.2$~dB was directly measured without any loss correction and circuit-noise correction. The measured squeezing levels with phase locking followed a Gaussian distribution whose standard deviation was 0.07~dB, and 99.7\% of the measured values fell within $\pm 3\sigma$ from the average.
Therefore, we concluded that the measurement error is approximately $0.2$~dB.
In a separate measurement, the shot-noise levels for various LO powers verified the linearity of the balanced homodyne detection system.

\begin{figure}[htbp]
    \centering\includegraphics[width=13.5cm]{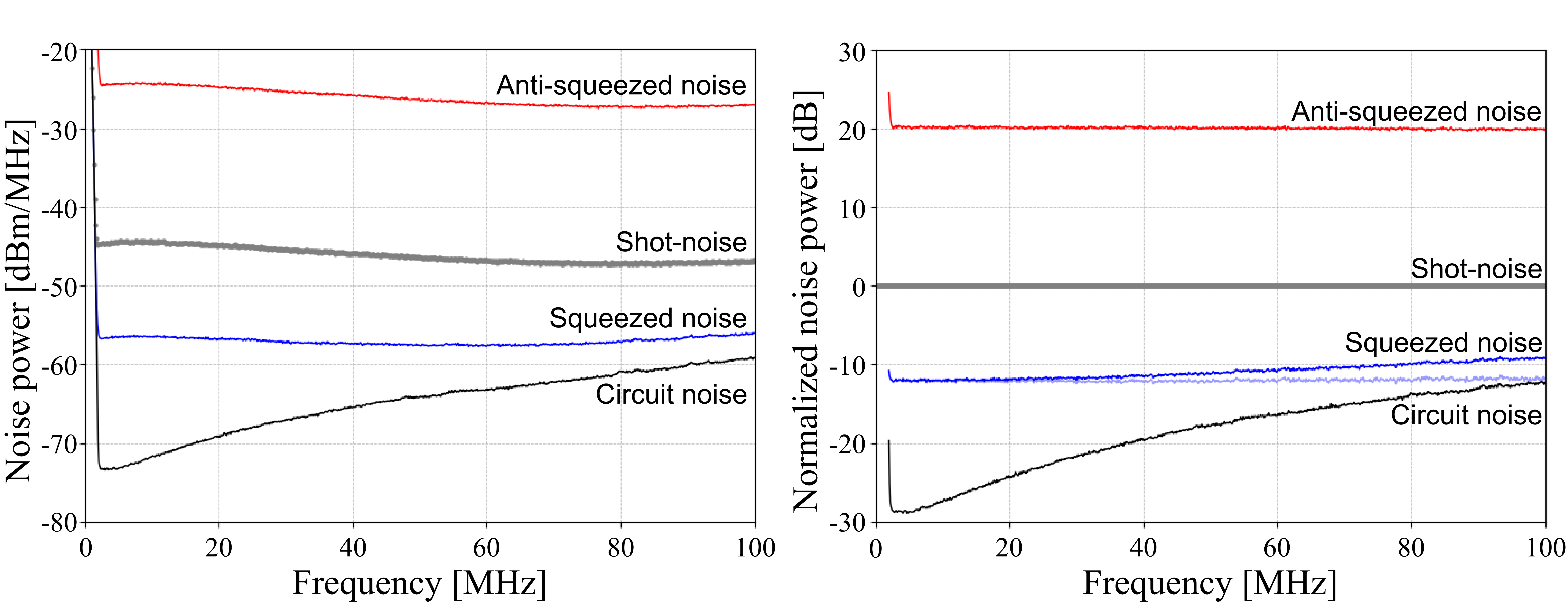}
    \caption{Left: Various noise levels at sideband frequencies up to 100~MHz. The pump power was 585~mW, measured immediately after the squeezer OPA. The resolution bandwidth was 1~MHz and the video bandwidth was 100~Hz. Right: Noise levels normalized to shot-noise. Light blue is the circuit-noise-corrected squeezed noise.}
    \label{fig:12dB_freq}
\end{figure}

Figure \ref{fig:12dB_freq} shows (normalized) frequency spectra of shot-noise, squeezed noise, anti-squeezed noise and circuit noise with pump power of 585~mW. The circuit-noise-corrected squeezing level is shown in light blue. A peak at 1~MHz can be seen as a result of the frequency-shifted probe beam. The squeezing level exceeding 12~dB can be seen around 10~MHz.
The measured squeezing level deteriorated slightly for higher sideband frequencies. This was due to reduced clearance above the circuit noise and the 35 MHz bandwidth of the homodyne photodiode. Correcting for circuit noise, the squeezing level near 100~MHz was 11.8~dB below shot-noise.

\section{Discussions}
\begin{figure}[htbp]
    \centering\includegraphics[width=13cm]{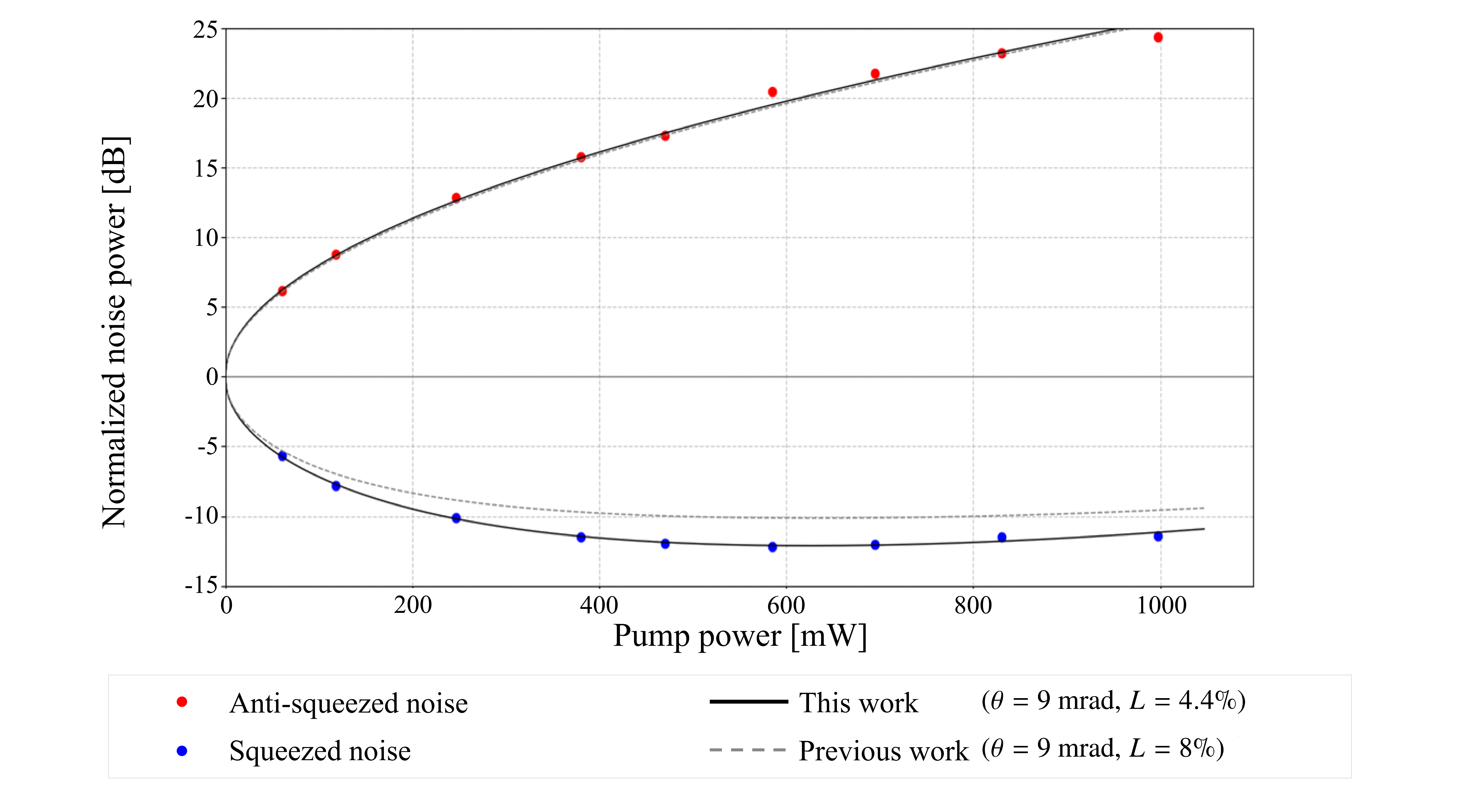}
    \caption{Squeezing and anti-squeezing levels for various pump powers, corrected for circuit noise. The solid curve is a fitting with parameters phase fluctuation of 9~mrad and loss of 4.4\%. Dotted line represents the theoretical curve with phase fluctuation and loss from our previous work \cite{Hirota10dB:26}. Each squeezed and anti-squeezed data point had measurement error of $\pm0.2$~dB.}
    \label{fig:squeeze_level_fitting}
\end{figure}

Loss $L$ from various experimental imperfections degrades the squeezing level $R_{-}$ and the anti-squeezing level $R_{+}$.
\begin{equation}
    R_{\pm} = L + (1-L) \exp(\pm2\sqrt{\alpha P})
\end{equation}
Here, $\alpha$ is the second harmonic generation (SHG) efficiency.
Imperfect phase lock results in a phase fluctuation between the squeezed light and the LO. The standard deviation of the phase fluctuation, denoted as $\tilde{\theta}$, degrades the squeezing level by adding noise proportional to the anti-squeezing level \cite{Zhang2003Phasenoise, Aoki:06, suzuki7.2dB2006};
\begin{equation}
    R'_{\pm} = R_{\pm} \cos^2\tilde{\theta} + R_{\mp} \sin^2\tilde{\theta}
\end{equation}
To analyze the loss and the phase noise, we measured the squeezed and anti-squeezed noise levels for various pump powers, as well as the shot-noise and the circuit noise. From these measurements, we subtracted the circuit noise and calculated the squeezing and the anti-squeezing levels (Fig. \ref{fig:squeeze_level_fitting}). By fitting the squeezing and anti-squeezing levels for various pump powers using the two equations above, we determine the SHG efficiency to be 876~\%/W, the phase fluctuation to be $9\pm1$~mrad, and the loss to be $4.4\pm0.4$\%. When compared to Ref. \cite{Hirota10dB:26}, we had the same amount of phase fluctuation, but a substantial reduction in loss, owing to the reduction in mode mismatch of the squeezed light and the LO.

We attribute the loss of 4.4\% to the following factors: loss inside the waveguide OPA, propagation loss, imperfect quantum efficiency of photodiodes, and mode mismatch between the squeezed light and the LO. First, following the calculation of effective loss inside a waveguide OPA described in \cite{Yamashima:25} and propagation loss of 0.1~dB/cm \cite{Kashiwazaki_6THz_0.1dBpercm}, we calculate a loss of around 2\%.
Second, the propagation loss of all optical elements in the path of the squeezed light, such as mirrors, lenses, and beam splitters, is measured as 1\%.
Third, the quantum efficiency of the photodiode was 99\%, corresponding to a loss of 1\%.
Finally, the remaining loss is due to the spatial mode mismatch between the squeezed light and the LO, which we estimate to be 0.4\%. Note, however, that this value is uncertain as the propagation loss within the waveguide and quantum efficiency of the photodiode were not measured to great accuracy.

It is important to note the physical mechanism behind this reduction in mode mismatch loss. While the waveguide used in this experiment was quasi-single-mode, fabrication and experimental imperfections can generate higher-order modes, making the generated squeezed light no longer single mode. Because free-space balanced homodyne detection inherently acts as a spatial filter, improving the squeezing level cannot be viewed merely as matching two pure single modes. Rather, by utilizing the squeezing level directly as the objective function, the BO algorithm molds the LO phase front to selectively overlap with the specific spatial mode (or superposition of modes) emitted by the OPA that exhibits the highest degree of squeezing. In this sense, the SLM performs spatial mode-selection alongside mode-matching, bypassing the strict single-mode requirements typically placed on waveguide OPA designs.

Further improvements for a higher squeezing level may include reducing intrinsic loss within the waveguide OPA. As this loss is a function of the length of the waveguide as well as the gain \cite{Yamashima:25}, reducing the length of the waveguide OPA while keeping the gain to a similar level is expected to decrease this loss. Other methods for improving the squeezing level include lowering the phase fluctuation. This includes matching the optical path lengths for the pump beam, probe beam, and the LO to mitigate laser noise \cite{Zhou_LaserNoise_2021, Cahillane_LaserNoise:2021}.

Finally, although the measured bandwidth of the squeezed light was limited by the homodyne detector, the method described in this paper does not impede the bandwidth of the squeezed light generated. Therefore, the squeezing level is estimated to have a THz-bandwidth, as reported in \cite{Kashiwazaki_6THz_0.1dBpercm}. Furthermore, provided that highly reflective, low-loss SLMs are developed, incorporating this double-reflection SLM configuration into the path of the signal beam could significantly enhance the detection efficiency of broadband real-time quadrature measurements of quantum states of light, such as those utilized for the generation and tomography of ultrafast non-Gaussian states \cite{Kawasaki_cat;2024}.

\section{Conclusion}
We generated squeezed light from a broadband waveguide OPA and measured a squeezing level of $12.1 \pm 0.2$~dB. This was realized by introducing an SLM controlled by machine learning in the path of the LO. Consequently, we greatly reduced the total loss to 4.4\%. This method of performing wavefront correction on the LO lowered the loss from mode mismatch to a level similar to that of the cavity-based systems \cite{15dB_OPO_PhysRevLett.117.110801, squeezing_at_1550nm_13.5dB}.
To further enhance the squeezing level, our next steps involve further reducing the internal propagation loss of the PPLN waveguide and implementing an even more stable phase-locking scheme to minimize phase jitter.
We believe that this achievement of 12~dB squeezing represents a major milestone toward the realization of universal quantum computers with THz-scale clock frequencies \cite{warit_2D_cluster, Takeda_and_Furusawa_2019_toward_FTUPQC}.

\begin{backmatter}
\bmsection{Funding}
Japan Science and Technology Agency (JPMJMS2064, JPMJPR2254). Japan Society for the Promotion of Science (24K01374).

\bmsection{Acknowledgment}
The authors thank Dr. Takeshi Umeki of NTT Device Technology Laboratories for providing the PPLN waveguides. The authors acknowledge supports from UTokyo Foundation and donations from Nichia Corporation of Japan. T.S. acknowledges financial support from The Forefront Physics and Mathematics Program to Drive Transformation (FoPM), a World-leading Innovative Graduate Study (WINGS) Program, the University of Tokyo. A part of this work was performed for Council for Science, Technology and Innovation (CSTI), Cross-ministerial Strategic Innovation Promotion Program (SIP), "Promoting Application of Advanced Quantum Technologies to Social Challenges" (Project management agency: QST).

\bmsection{Disclosures}
The authors declare no conflicts of interest.

\bmsection{Data Availability}
The data that support the findings of this work are available from the corresponding author upon reasonable request.
\end{backmatter}

\bibliography{sample}

\end{document}